\documentclass[aps,prb,twocolumn,epsf,showpacs,amsmath,multicol]{revtex4}
\usepackage{graphicx}
\def\be{\begin{equation}}
\def\ee{\end{equation}}
\def\bea{\begin{eqnarray}}
\def\eea{\end{eqnarray}}

\def\a{\alpha}
\def\b{\beta}
\def\g{\gamma}

\def\d{\delta}
\def\D{\Delta}
\def \e{\varepsilon}
\def\h{\hbar}

\def\Tr{\mbox{tr}}
\def\S{{\cal S}}
\def\l{\lambda}
\def\L{\openone}
\def\m{\mu}
\def\o{\omega}
\def\O{\Omega}
\def\V{\hat V}

\def\Bolts {k_{\rm B}}

\def\dwell{\tau_{\rm d}}
\begin{document}
\title{Photon-assisted electron-hole shot noise in multi-terminal conductors}
\date{\today}
\author{Valentin S. Rychkov}
\email{valentin.rytchkov@physics.unige.ch}
\author{Mikhail L. Polianski}
\author{Markus B\"uttiker}
\affiliation{D\'epartement de Physique Th\'eorique, Universit\'e
de Gen\`eve, CH-1211 Gen\`eve 4, Switzerland}
% 73.23.-b    Electronic transport in mesoscopic systems
%73.21.La    Quantum dots
%73.50.Pz    Photoconduction and photovoltaic effects
%73.50.Td    Noise processes and phenomena
%73.23.Hk    Coulomb blockade; single-electron tunneling
\begin{abstract}
Motivated by a recent experiment by L.-H. Reydellet \emph{et al.},
Phys. Rev. Lett. \textbf{90}, 176803 (2003), we discuss an
interpretation of photon-assisted shot noise in mesoscopic
multiprobe conductors in terms of electron-hole pair excitations.
AC-voltages are applied to the contacts of the sample. Of interest
are correlations resulting from the fact that electrons and holes
are generated in pairs. We show that with two out-of-phase
ac-potentials of equal magnitude and frequency, applied to
different contacts, it is possible to trace out the Hanbury Brown
Twiss exchange interference correlations in a four probe
conductor. We calculate the distribution of Hanbury Brown Twiss
phases for a four-probe single channel chaotic dot.

\end{abstract}
 \maketitle
\section{Introduction}
Theoretical and experimental investigations of the current and
charge noise properties of small conductors are an important
frontier of mesoscopic physics. The aim is to analyze noise as an
additional source of information on the quantum statistical
properties of small conductors. Interest in quantum communication
and quantum computation have further drawn attention to this
subject. We refer the interested reader to
reviews,\cite{review1,review2} a conference book with extended
articles on the subject,\cite{nazarov} a special issue of a
journal,\cite{special} and to original work on shot noise in
mesoscopic conductors.
\cite{khlus,lesovik,buttiker90,buttiker91,birk,reznikov,kumar,oberholzerprl}
Investigations have predominantly considered samples subject to
stationary applied voltages which drive a dc-current through the
sample. The work reported here is motivated by a recent experiment
by Reydellet \emph{et al.} [\onlinecite{glattli}] which applies
only an ac-voltage to a contact of a two-terminal sample. There is
no dc-current linear in voltage. However, the ac-voltage leads to
the generation of electron-hole pairs and the members of this pair
are subsequently transmitted through the sample or reflected back
into the excited contact.

For several reasons, it is of interest to analyze the noise in
terms of electron-hole excitations. First, if an ac-voltage is
applied to a conductor that is otherwise at equilibrium,
electron-hole pairs are in fact the natural elementary excitations
of the system. \cite{mosk} This is in contrast with most of the
literature on shot noise which even in the presence of
ac-excitations uses an electron picture only. The understanding
and interpretation of the results can differ dramatically
depending on whether one relies on an all electron picture or on
an electron-hole picture. A second reason to investigate
electron-hole pair generation is that recently it was realized
that such pairs are a source of entanglement, similar to the
optical process in which a pair of photons with entangled
polarization state is generated through down
conversion.\cite{bee1} Indeed an electron-hole pair creation event
leads to an electron and hole who's spins are
entangled.\cite{bee1} However, it is not only the spin degree
which can be useful, but electron-hole sources can be used to
generate orbital \cite{sam1} quasi-particle entanglement. This
leads to simple and controllable geometries \cite{sam2} which
permit the investigation of entanglement in electrical conductors
without the need of superconducting/semiconductor/ferromagnetic
hybrid structures. Indeed the dynamic generation of electron-hole
pairs through the periodic modulation of potentials in the {\it
interior} of conductors \cite{sam3,bee2} or through the
application of pulses \cite{llb} has recently been discussed.
\begin{figure}
\includegraphics[height=5cm]{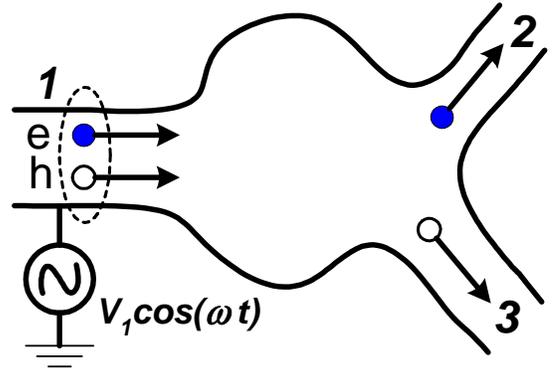}
\caption{An ac-bias voltage $V_1\cos\o t$ in lead $1$ excites
correlated ($e$) electron - ($h$) hole pairs.  In the
zero-temperature limit, the zero-frequency current
cross-correlation between the second and third lead is generated
only by pairs which are split, one carrier sent into lead $2$ and
one carrier into lead $3$. The current-correlation is a direct
measure of the coincident generation of electron and hole in a
pair.} \label{fig1}
\end{figure}

Fig. \ref{fig1} shows the type of structures which we analyze in
this work. Electron-hole pairs are excited in contact $1$. The
rate with which carriers are excited is much smaller than the
inverse time taken by an electron or hole to be transmitted
through the conductor. Fig. \ref{fig1} depicts a particular
scattering event in which the pair is split and the electron
leaves the conductor through contact $2$ and the hole leaves the
conductor through contact $3$. An essential point of our work is
the fact the correlation of currents between contacts $2$ and $3$
is entirely determined by scattering events in which an
electron-hole pair is split. Other scattering processes, for
instance one in which only one carrier of the pair leaves the
conductor through contact $2$ and the other is scattered back into
contact $1$, obviously give no contribution to the current
correlation. Thus in the limit of rare excitations of
electron-hole pairs, the current-correlation between contact $2$
and $3$ is a direct measurement of the coincident creation of an
electron and hole.

The term 'photon-assisted transport' is applied here in the
following sense: For weak coupling of the photon density with the
electrons in the lead, the photon field appears in the many-body
electron Hamiltonian as a weak periodic potential $V(t)$. We can
neglect the feedback of the weak currents on the photon source.
Thus in the following, we consider all photon sources as weak
ac-bias potentials applied to the leads of the conductor.
\cite{tuck}

Photon-assisted current in quantum dots have been experimentally
investigated in Ref. [\onlinecite{accurrent}]. But only a few
experiments have thus far investigated the noise properties of
photon-assisted transport. Following the theoretical work of
Lesovik and Levitov,\cite{LL} Schoelkopf \emph{et al}.
\cite{schoel} illuminated a normal diffusive conductor and found
features in the zero-frequency shot noise at the photon energies
$V = nh\nu /e$. Theoretical work on hybrid-superconducting system
by Lesovik \emph{et al}. \cite{les99} lead similarly to the
experimental observation of features at energies $V = nh\nu /2e$
by Kozhevnikov \emph{et al}. \cite{kozh} In both of these
experiments a dc-voltage was applied in addition to the radiation.
A similar regime for the systems in the fractional quantum Hall
state is discussed in Ref. [\onlinecite{tmart}]. The theoretical
works \cite{LL,les99} consider photon-assisted transport purely as
a particle transport problem. However, despite the fact that one
measures a dc-current or a zero-frequency noise spectrum, the
radiation actually also generates a charge and current response at
the excitation frequency. The important role of displacement
currents and their rectification effects have been emphasized in
the work of Pedersen and Buttiker.\cite{pedersen} In the work
presented here we will also only consider non-interacting
particles and derive expressions for electron and hole currents
and their correlations. The role of Coulomb interactions,
inelastic scattering and dephasing on the photon assisted noise of
electron-hole pairs is treated in the Ref. [\onlinecite{poli}].

In contrast, to the above mentioned experiments, the recent shot
noise measurements by Reydellet \emph{et al.}
[\onlinecite{glattli}] report results for the case that only an
ac-voltage is applied to the leads. Reydellet \emph{et al.}
[\onlinecite{glattli}] also discuss their results in terms of
electron-hole excitations assuming that the electrons and holes
generate a partition noise independently from each other and add
incoherently. As indicated above our interest is in the {\it
correlated} nature of the electron-hole processes and the
manifestation of this correlation in shot-noise spectra. We first
derive general expressions for electron and hole currents and
subsequently apply them to two, three and four terminal
conductors. Of particular interest is an arrangement which uses
two out-of-phase ac-voltages of the same magnitude and frequency.
The phase difference between the applied voltages provides an
additional degree of control. It turns out that the exchange
contribution to the current correlations
\cite{buttiker90,buttiker91,VLB,sam2} (the Hanbury Brown Twiss
(HBT) correlations \cite{hbt}) are sensitive to the phase
difference between the applied voltages. The HBT-correlations
reflect indistinguishable two-particle processes in the shot
noise. \cite{sam2} Sweeping the phase shift allows to maximize
(minimize) shot-noise correlations. The positions of these extrema
yield information about combinations of phases of the scattering
matrix, which we call the HBT-phase. If the phase-difference of
the ac-potentials is set to the HBT-phase, the correlation is
maximal. This permits to explore the statistical properties of the
HBT-phase. We analyze the distribution of the HBT-phase for a
chaotic dot coupled to single channel leads.

The additional degree of control generated by two potentials which
are out of phase has its analog in quantum pumping where a
scatterer is modulated with two out-of-phase parameters.
\cite{B_p,Switkes,Avron,VAA,PVB}$^,$ \cite{mosk} Adiabatic quantum
pumping uses in an essential way this additional degree of
control. The difference is that here we deal with potentials
applied to contacts and we analyze a situation in which the
frequency to voltage ratio is large.

The paper is organized as follows: In Sec. \ref{sec:system} we
define the geometry of the mesoscopic conductors, present the
model assumptions and the theoretical approach used. Sec.
\ref{sec:eh} describes the electron-hole picture for
photon-excited multiprobe conductors and presents sample-specific
results for the shot-noise spectra. Sec. \ref{sec:phase} discusses
an arrangement to measure the Hanbury Brown Twiss exchange
interference correlations. We conclude in Sec.
\ref{sec:conclusions}. Appendix \ref{sec:exchange} gives the
quantum state which leads to the HBT-exchange interference
correlations, and Appendix \ref{sec:Poisson} presents a
probability theory argument to show the existence of electron-hole
correlations.

%-----------------------------------------------------------------------------
%------------------------------------------------------------------------------
\section{Scattering approach}\label{sec:system}

We consider a multi-terminal mesoscopic conductor connected to
metallic contacts. The leads are subject to time-dependent
voltages. We briefly recall the main aspects of the scattering
approach and introduce the basic notations. We follow Refs.
\cite{pedersen,Hartree} who distinguish external potentials
(applied to leads) from internal potentials (generated in the
interior of the conductor). This distinction is particularly
useful to a theory formulated in terms of the scattering matrix:
external potentials leave the scattering matrix invariant whereas
internal potentials lead to emission and absorption of energy
inside the mesoscopic conductor. A realistic description of
time-dependent transport involves both external and internal
oscillating potentials. Here we focus on the effect of external
potentials, keeping the internal potential fixed.

Metallic reservoirs are considered as emitters (absorbers) of
charge carriers incident on (exiting from)  the mesoscopic
conductor. Motion of electrons in the leads can be described by
their energy, transverse mode number and direction of momenta.
Thus we introduce operators of incoming and outgoing states: ${\bf
a}_\l(\e_i)$ and ${\bf b}_\l(\e_i)$ are vectors of operators
$a_{\l n}(\e_i)$ and $b_{\l n}(\e_i)$. The operator $a_{\l
n}(\e_i)$ annihilates  a carrier incident in the $\l$-th lead with
energy $\e_i$ in the $n$-th transverse channel. The operator
$b_{\l n}(\e_i)$ annihilates an outgoing carrier in the $\l$-th
lead with energy $\e_i$ in the $n$-th transverse channel. Here and
in the following greek indices run over all contacts of the
conductor. For a stationary scatterer the connection between
incoming and outgoing carriers is governed by scattering matrices
which depend on one energy only. We consider the case of spinless
electrons. The scattering matrix transforms the operators ${\bf
a}_\l(\e_i)$ into ${\bf b}_\l(\e_i)$ according to ${\bf
b}_\a(\e)=\sum_\b \S_{\a\b}(\e){\bf a}_\b(\e)$. In a multichannel
conductor, the matrix $\S_{\a\b}(\e)$ is
 a sub-block of the scattering matrix $\S$ for scattering from lead $\b$
($N_\b$ channels) to lead $\a$ ($N_\a$ channels) with energy $\e$.
It has dimensions $N_\a\times N_\b$.

The current operator in contact $\l$ is \cite{buttiker90}

\begin{eqnarray}\label{eq:current1}  I_\l(t) &=& \frac{e}{2\pi\h}\int\int d\e_1
d\e_2e^{i(\e_1-\e_2)t/\h}
 \nonumber \\&\times& \sum_{\a\b}{\bf
a}^\dag_\a(\e_1){\bf A}_{\a\b}(\l,\e_1,\e_2) {\bf a}_\b(\e_2)
.\end{eqnarray}
Using the projector
matrix $\openone_\l$ which is a unit matrix of size $N_\l \times
N_\l$ in the $\l$-th lead, we introduce the current matrix \cite{buttiker90}

\begin{equation}\label{current_matrix} {\bf A}(\l,\e_1,\e_2)=\L_\l-\S^\dag(\e_1)
\L_\l \S(\e_2). \end{equation}
In the following we use the abbreviated notation ${\bf A}(\l,\e)$
if the energies coincide and ${\bf A}(\l)$ if the energy
dependence is unimportant. Electrons obey Fermi statistics. In
order to transfer charge across the mesoscopic conductor the
incoming charge state has to be filled and at the same time the
outgoing state has to be open. A dc-voltage applied across the
sample opens an energy window where both conditions are fulfilled
and transport is possible.

Consider now a time-dependent potential $eV_\a(t)=eV_\a\cos(\o
t+\phi_a)$ applied to the $\a$-th lead. This potential can be
absorbed in the phase of the wave function. The single particle
wave function in the presence of the perturbation is: $\psi_{\a
n}(\e,t)=\phi_{\a n}(\e,t)\exp\{-ieV_\a\sin(\o t+\phi_a)/\h\o)\}$.
Here $\phi_{\a n}(\e,t)$ is the stationary wave function
describing incoming (outgoing) carriers in the $n$-th transverse
channel with energy $\e$. This wave function can be expressed as a
series in Bessel functions:

\begin{equation} \label{wave-function-time-dep} \psi_{\a n}(\e,t)=\phi_{\a
n}(\e,t) \sum_l J_{l}\left(\frac{eV_\a}{\h\o}\right)e^{-il(\o
t+\phi_\a)}.
\end{equation}
We see, that wave function $\psi_{\a n}(\e,t)$ in Eq.
(\ref{wave-function-time-dep}) has the same coordinate dependence
as $\phi_{\a n}(\e,t)$ but in energy space it is a superposition
of sideband states with amplitudes
$J_{l}(eV_\a/\h\o)\exp(-il\phi_\a)$ and energies $\e-l\h\o$.

We follow Refs. [\onlinecite{Hartree,pedersen}] and assume that
the oscillating potential exists in a region of the lead between
the reservoir and the conductor. The potential vanishes as we
approach the conductor. Matching wave functions in the regions
with and without oscillating potential leads to

\begin{eqnarray}
\label{eq:operator_connection} a_{\a n}(\e)=\sum_l a'_{\a
n}(\e-l\h\omega)J_{l}\left(\frac{eV_\a}{\h\o}\right)e^{-il\phi_\a}.
\end{eqnarray}
Here and in the following $a_{\a n}(\e)$ are operators
corresponding to annihilation of carrier incident on the
conductor, $a'_{\a n}(\e-l\h\omega)$ are operators corresponding
to annihilation of carriers in the reservoir. This expression has
an accuracy of order $\h\o/\e_F$. We neglect corrections
\cite{Hartree} which arise from the difference of momenta of the
particles with different sideband energies $\e$ and $\e-l\h\o$.
The reservoir is at thermal equilibrium and the statistical
average of the operators $({\bf a'}_\a)^{\dag}$ and ${\bf a}_\b'$
corresponds to the Fermi distribution of electrons in the absence
of the time-dependent voltage: $\langle({\bf
a'}_\a)^{\dag}(\e_1){\bf
a}_\b'(\e_2)\rangle=\d_{\a\b}\d(\e_1-\e_2)f_\a(\e_1)$.

Let us next discuss the noise. At zero frequency, the noise spectral
density is defined as follows:

\begin{eqnarray}\label{sn1}
S_{\l\mu}&=&\lim_{T\rightarrow\infty}\frac{1}{T}\int_{0}^{T}dt\int_{-\infty}^{\infty}d\tau
\langle\D I_\l(t+\tau)\D I_\mu(t)\rangle. \end{eqnarray}
Here angular brackets denote the quantum mechanical and
statistical average and $\D I_\l(t)= I_\l(t)-\langle
I_\l(t)\rangle$. For time independent voltages the current
correlator depends only on the time difference, hence the integral
over time $t$ in Eq. (\ref{sn1}) is trivial and leads to the
stationary state expression for shot noise.\cite{review1} In
general, in the case of  a time-dependent perturbation, the
current correlator depends on both times $t$ and $\tau$. One can
think about $\tau$ as the time difference and $t$ as the moment at
which the measurement starts. In the steady state situation, a
shift of the time $t$ by an integer number of periods changes
nothing. This is why the averaging over time $t$ can be performed
only over one period of the perturbation $\displaystyle
T_0=2\pi/\o$.
\begin{eqnarray}\label{sn3}
S_{\l\mu}=\frac{1}{T_0}\int_{0}^{T_0}dt\int_{-\infty}^{\infty}
d\tau \langle\D I_\l(t+\tau)\D I_\mu(t)\rangle. \end{eqnarray}
Below, we will write this current noise in terms of pure electron
and pure hole noise contributions and their correlations.

\section{Electron-hole picture of noise}\label{sec:eh}

In the presence of ac-potentials it is appropriate to consider
excitations away from the global equilibrium state and these
excitations are electron-hole pairs. Our goal, motivated by the
recent experiment of Reydellet \emph{et al}.
[\onlinecite{glattli}], is to develop an electron-hole description
of photon-assisted shot noise. The creation of an electron-hole
pair is a correlated process. These correlations have been used in
proposals for the generation and detection of orbital
quasi-particle entanglement. \cite{sam1,bee1,sam2} Therefore the
manifestations of electron-hole correlations in the noise
properties are of particular interest.

To distinguish between different contributions to the shot noise
we express the incident states in terms of electron and hole
operators,

\begin{eqnarray}\label{particle-hole}
  {\bf e}_\a(\e)= {\bf a}_\a(\e)\theta_e(\e)\ ,\ {\bf h}_\a(\e)={\bf
  a}_\a^\dag(\e)\theta_h(\e),
\end{eqnarray} where $\theta_e(\e)=\theta(\e)$ and $\theta_h(\e)=\theta(-\e)$
are step functions for electrons and holes. Note that ${\bf
h}_\a(\e)$ is a transposed vector (row). Equation (\ref{sn3}) for
the shot noise can be converted to energy space. Using the Fourier
transform of the current operator $I_\l(t)=\int d\e d\O\exp(i\O
t)I_\l(\e+\O,\e)$ and recalling the fact that the condition
(\ref{eq:operator_connection}) implies $\e_1-\e_2 = l\h\o$ for the
current operators we obtain the following result for the
zero-frequency shot noise:

\begin{eqnarray} \label{shot_noise_step3} S_{\l\mu}&=&2\pi\h\int \int d\e_1d\e_2
\langle\Delta  I_\lambda(\e_1)\Delta I_\mu(\e_2)\rangle.
\end{eqnarray}
The current operator $I_\l(\e)\equiv I_\lambda(\e,\e) $ at equal
energies can be represented as a sum of electron and hole currents
$I_\l(\e)=I_{\l}^e(\e)+I_{\l}^h(\e)$.  Using Eq.
(\ref{particle-hole}) and Fourier transform of Eq.
(\ref{eq:current1}) we obtain:
\begin{eqnarray}
\label{eq:ceh1} I_\l^{e}(\e)&=&\frac{e}{2\pi\h}\sum_{\a\b}{\bf
e}^\dag_\a(\e){\bf A}_{\a\b}(\l,\e){\bf e}_\b(\e),\\
\label{eq:ceh2}I_\l^{h}(\e)&=&-\frac{e}{2\pi\h}\sum_{\a\b}{\bf
h}^*_\a(\e){\bf A}^T_{\a\b}(\l,\e){\bf h}^T_\b(\e),
\end{eqnarray}where the electron charge is $e=-|e|$.
It is now interesting to calculate explicitly the correlations
between the same or different types of particles:
\begin{eqnarray}
\label{shot_noise_eh1} S_{\l\mu}&=&\sum_{ij=eh} S^{ij}_{\l\mu}=
2\pi \h\sum_{ij=eh}\int\int d\e_1d\e_2 \langle \Delta
I^{i}_\lambda(\e_1) \Delta I^{j}_\mu(\e_2)\rangle.\nonumber
\end{eqnarray}
A non-zero answer for $S^{eh}_{\l\m}$ will show us the presence of
intrinsic correlations between electrons and holes.  Using the
expressions for the currents (\ref{eq:ceh1}), (\ref{eq:ceh2}) and
Eqs. (\ref{eq:operator_connection}) and (\ref{particle-hole}) we
find the correlations of the electron and the hole currents:
\begin{eqnarray}\label{eq:ij}
S_{\lambda\mu}^{ij}&=&\frac{e^2}{2\pi\h}\sum_{klm\alpha\beta}\int\int
d\e_1d\e_2\delta(\e_1-\e_2+m\h\omega) \nonumber \\
&\times&
J_{k}\left(\frac{eV_\a}{\h\o}\right)J_{k+m}\left(\frac{eV_\a}{\h\o}\right)J_{l+m}\left(\frac{eV_\b}{\h\o}\right)J_{l}\left(\frac{eV_\b}{\h\o}\right)\nonumber\\
&\times &e^{im(\phi_\b-\phi_\a)}f_\a(\e_2-k\h\omega)(1- f_\b
(\e_2-l\h\omega))\nonumber\\
&\times&\Tr\left({\bf A}_{\alpha\beta}(\lambda,\e_1){\bf
A}_{\beta\alpha}(\mu,\e_2)\right)\theta_i(\e_1)\theta_j(\e_2).
\end{eqnarray}
Here $i=e,h$ stands for electron or hole. For example, in the
limit of in-phase applied voltages the sum of Eq. (\ref{eq:ij})
over electron and hole indices gives the result which for a
two-probe sample with energy-independent scattering matrix can be
found in Ref. [\onlinecite{LL}] and for a multi-probe conductor is
given in Ref. [\onlinecite{pedersen}].

To probe correlations of electron-hole pairs we impose two
conditions. First, the ac-potentials are taken to be weak enough
in order to exclude processes of multiple absorbtion or emission
of photons which implies $eV_\a\ll\h\o$. As a consequence,
electron-hole pairs are generated one by one. In this case one can
prove that the many body wave function of the system will
correspond to incoming single electron-hole pairs (see
[\onlinecite{sam3,sam1}] and Appendix \ref{sec:exchange}). Second,
electrons and holes have different energies (the typical energy
scale of an electron-hole pair is $\h\o$). We neglect the energy
dependence of scattering matrices on the scale of $\h\o$. The
energy dependence of the scattering matrix starts to play a role
when the frequency becomes comparable to the inverse of the dwell
time $\dwell$ such that $\o\dwell\sim1$. When screening is taken
into account the frequency must even be comparable to the inverse
charge relaxation time. \cite{poli} As a consequence of these
conditions, the correlations of $ee$ and $hh$, as well as $eh$ and
$he$ correlations, are identical. Neglecting corrections of the
order of $(eV/\h\omega)^4$, we find the correlations between
electron and hole currents:

\begin{eqnarray}\label{eq:ee_eh}
S_{\lambda\mu}^{ee}&=& \frac{e^2\o}{2\pi}
\Tr\,{\bf A}(\mu){\cal P}{\bf A}(\l),\\
S_{\lambda\mu}^{eh}&=&-\frac{e^2\o}{2\pi}\Tr\,{\bf
A}(\mu)\sqrt{\cal P}e^{i\phi}{\bf A}(\l)\sqrt{\cal P}e^{-i\phi}.
\end{eqnarray}

Similarly to the experimental work\cite{glattli} we introduce the
probability to create an electron-hole pair in the $\a$-th channel,
${\cal P}_\a=(eV_\a)^2/(2\h\o)^2$ and diagonal matrices ${\cal
P}=\mbox{ diag}({\cal P}_1,...,{\cal P}_N)$ and $\phi=\mbox{diag
}(\phi_1,...,\phi_N)$ are defined by the amplitudes and phases of
applied voltages. We next consider a number of different set-ups.

Consider a multi-lead conductor and consider an alternating
voltage applied to only one lead, say lead $1$. The probability to
create an electron-hole pair is ${\cal P}_1 =(eV_1/2\h\o)^2$. All
other leads are grounded. The auto-correlations and
cross-correlations are
\begin{eqnarray}\label{eq:electron-hole-noise-answer}
S^{ee}_{\l\mu}&=& \frac{e^2\o}{2\pi}{\cal P}_1
\Tr\,{\bf A}(\m)\openone_1{\bf A}(\l), \\
S^{eh}_{\l\mu}&=& -\frac{e^2\o}{2\pi}{\cal P}_1\Tr\,{\bf
A}(\m)\openone_1{\bf A}(\l)\openone_1.
\end{eqnarray}

Next consider the 2-terminal case. Charge current conservation and
particle (electron and hole) current conservation imply that the
auto-correlations and the cross-correlations are equal in magnitude
and differ only by a sign. Thus it is sufficient to give the
auto-correlations,

\begin{eqnarray}
\label{eq:2term_corr} S^{ee}_{11}&=& \frac{e^2\o}{2\pi}{\cal P}_1
\sum_n{\cal T}_n,\,\,\, S^{eh}_{11}= -\frac{e^2\o}{2\pi}{\cal P}_1
\sum_n{\cal T}_n^2.
\end{eqnarray}
Here ${\cal T}_n$ is transmission probability of the $n-th$
eigen-channel, \emph{i.e.} an eigenvalue of
$\S^{\dagger}\openone_2\S\openone_1$. Summing up all four terms
gives a shot noise of the total measured charge current
proportional to $\sum_{n} {\cal T}_n (1-{\cal T}_n )$. However,
unlike the case of a dc-biased conductor, where the ${\cal T}_n
(1-{\cal T}_n )$ is proportional to the quantum partition noise of
a \emph{fully filled} incident channel,\cite{lesovik,buttiker90}
here ${\cal T}_n (1-{\cal T}_n )$ has a completely different
origin. The auto-correlations of the electron $S^{ee}_{11}$ and
hole noise $S^{hh}_{11}$ are simply proportional to ${\cal T}_n$
reflecting Poissonian shot noise of a nearly \emph{empty channel}
of incident particles. Electron and hole particle currents are
correlated and it is this two particle correlation which is
proportional to ${\cal T}_n^2$. The existence of electron-hole
correlations can also be proved using a simple probability theory
argument, which is demonstrated in Appendix \ref{sec:Poisson}.

In order to make a direct measurement of electron-hole
correlations we now analyze cross-correlations in three terminal
mesoscopic conductors (see Fig. \ref{fig1}). For a weak
ac-perturbation electron-hole pairs are rarely injected into the
mesoscopic conductor. If we measure the current-correlation at the
grounded leads, then if electron and hole exit through the same
lead or if one of the particles of the pair is reflected back then
their contribution to the cross-correlation is zero. Thus current
cross-correlations are determined only by electron-hole pairs for
which one of the particles exits through contact $2$ and the other
through contact $3$ (see Fig. \ref{fig1}). In fact a
cross-correlation measurement is a coincidence measurement run
over long times. \cite{sam1}

For a three terminal conductor we find for the particle
cross-correlations,

\begin{eqnarray}\label{three_terminal_correlations}
S^{eh}_{23}= -\frac{e^2\o}{2\pi}{\cal P}_1\sum_n\frac {{\cal
T}_{21n}{\cal T}_{31n}}{2},\,\, \,S^{ee}_{23} =0 .
 \end{eqnarray}
Here ${\cal T}_{21n}$ and ${\cal T}_{31n}$ are transmission
eigenvalues of $\S^{\dagger}\openone_2\S\openone_1$ and
$\S^{\dagger}\openone_3\S\openone_1$. This demonstrates that
 the
current-cross-correlations are a direct measure of electron-hole
correlations.

The considerations made above are valid at zero-temperature. It is
thus important to consider the effect of thermal noise. After all,
thermal noise can be viewed as another mechanism which generates
electron-hole pairs. It should not matter how exactly
electron-hole pairs are created. From the above consideration we
see that electron-hole correlations are of second order in the
transmission probability both for auto-correlations and for
cross-correlations. We know from FDT that equilibrium noise is
proportional to conductance and hence it is proportional to
transmission probability. Consequently we do not expect any
electron-hole correlations in thermal equilibrium.

Equilibrium noise is given by Eq. (\ref{eq:ij}); only the terms
$k=l=m=0$ contribute at $V=0$. In Eq. (\ref{eq:ij}) the expression
$\theta_e(\e_1)\theta_h(\e_2)\delta(\e_1-\e_2)\equiv 0$, so the
only contribution to the equilibrium noise comes from
electron-electron and hole-hole correlations, both are
proportional to the dc-conductance matrix
$G_{\l\mu}=(e^2/h)(N_\l\d_{\l\mu}-\Tr\
\S^\dag\openone_\l\S\openone_\m)$:
\begin{eqnarray} {
S}_{\lambda\mu}^{ee}&=& 2\Bolts TG_{\lambda \mu},\,\,\, {
S}_{\lambda\mu}^{eh} =0 \label{thermal}\,. \end{eqnarray}

Notice that a system at thermal equilibrium does not exhibit
electron-hole correlations, as expected, whereas electron-electron
and hole-hole correlations do of course exist. Also Eq.
(\ref{thermal}) shows that the thermal contribution to the noise
can be neglected if $\Bolts T\ll(eV)^2/(\h\o)$.

\section{Hanbury Brown Twiss phase}\label{sec:phase}

If currents are injected from two or more contacts there are
contributions to current-cross-correlations $S_{34}$ which can not
be expressed in terms of transmission probabilities.
\cite{buttiker91} The cross-correlations also depend on terms
which contain products of four scattering matrices of the type
\cite{buttiker90,buttiker91,VLB}
$\S^\dag_{13}\S_{32}\S^\dag_{24}\S_{41}$. In such a product none
of the scattering matrices is the hermitian conjugate of the
other. As a consequence the cross-correlation depends on the {\it
relative} phase of scattering matrix elements. The physical origin
of these terms is the quantum mechanical indistinguishability of
particles. A particle from source contact $1$ can be transmitted
to either $3$ or $4$ and is indistinguishable form a particle
injected through contact $2$ (see Fig \ref{fig3}). The quantum
state is given explicitly in Appendix \ref{sec:exchange}.
\begin{figure}
\includegraphics[height=5cm]{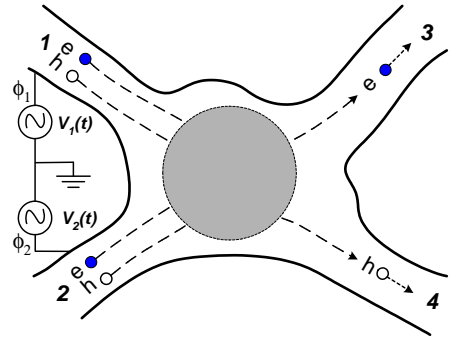}
\caption{Electron-hole pairs originating from the lead $1$ or $2$
contribute to cross-correlations measured in $3$ and $4$.
Indistinguishability of the sources leads to the dependence of the
shot noise $S_{34}$ on the relative phase $\Delta\phi$ of the
voltage sources } \label{fig3}
\end{figure}

This exchange interference effect reflects the fact that two
particles can be simultaneously in the conductor.  It has long
been proposed that this can be used to generate a two-particle
Aharonov-Bohm effect which exists only in the current correlations
while under the same conditions conductance exhibits no
Aharonov-Bohm effect at all. \cite{buttiker91} Recently a geometry
which demonstrates this explicitly has been proposed and analyzed.
\cite{sam2} The magnitude of the current cross-correlations is
shown to depend on the phase
\begin{equation}
\label{eq:HBT-phase}
\chi=\arg\left(\S^\dag_{13}\S_{32}\S^\dag_{24}\S_{41}\right),
\end{equation}
which we call Hanbury Brown Twiss (HBT) phase.

For simplicity, let us consider a four terminal mesoscopic
conductor with one mode contacts. AC-voltages of equal magnitude
are applied to leads $1$ and $2$ and generate electron-hole pairs
with probability ${\cal P}=(eV/2\h\o)^2$. The phase difference
between the oscillating potentials $V_1(t)$ and $V_2(t)$ is
$\Delta\phi$. Using (\ref{eq:ee_eh}) we find the charge current
correlation at contact $3$ and $4$:

\begin{eqnarray}\label{eq:cross_corr_phase_shift}
S(\Delta\phi)=-\frac{e^2\o}{2\pi}{\cal P}
\left|\S^\dag_{13}\S_{41}e^{-i\Delta\phi}+\S^\dag_{23}\S_{42}
\right|^2. \end{eqnarray}
Once created in the lead $1$ or $2$ an electron-hole pair
contributes to the shot noise only if it is split to the leads $3$
and $4$. The relative phase of pairs injected from $1$ and $2$ is
$\Delta\phi$. Indistinguishability of particles emitted from the
sources $1$ and $2$ leads to the phase dependence of the shot
noise. From Eq. (\ref{eq:cross_corr_phase_shift}) we calculate the
phase shift $\Delta\phi$ which corresponds to the maximum or
minimum in the shot noise, $\phi_\pm$ and the extremal values of
the cross-correlations $S^\pm$:
\begin{eqnarray}
\Delta\phi_+ &=&
\arg\left(\S^\dag_{13}\S_{32}\S^\dag_{24}\S_{41}\right)\label{eq:phase},\,
\Delta\phi_-=\Delta\phi_+ +\pi,\\
S^\pm &=&-\frac{e^2\o}{2\pi}{\cal
P}\left(|\S^\dag_{13}||\S_{41}|\pm|\S^\dag_{23}||\S_{42}|\right)^2.
\end{eqnarray}
Notice, that when the phase shift becomes equal to the HBT-phase
of Eq. (\ref{eq:HBT-phase}), $\Delta\phi=\chi$, the shot noise
reaches its maximum value $S^+$. For a dc-biased 4 terminal
conductor one can extract information about
$Re(\S^\dag_{13}\S_{32}\S^\dag_{24}\S_{41})$ using results of
three different measurements.\cite{buttiker91,VLB} In contrast,
the ac-perturbation gives an additional degree of freedom to vary
the phase shift between the voltages. Periodic dependence of the
noise on the phase shift allows us to reproduce the whole function
$S(\Delta\phi)$ using measurements at three different values of
the phase shift $\Delta\phi$. Maximal $S^+$ and minimal $S^-$
values of the shot noise, as well as the corresponding phase
shifts $\Delta\phi_{\pm}$, yield not only the real part of
$\S^\dag_{13}\S_{32}\S^\dag_{24}\S_{41}$ but also the full
exchange interference correlation, \emph{i.e.} its imaginary part.

We next investigate the statistical properties of the HBT-phase of
four-terminal conductors. First, it is useful to mention that
especially simple structures like a Mach-Zehnder (MZ)
interferometer which was recently realized in  a 2DEG \cite{mz}
exhibit only a trivial HBT-phase. A Mach-Zehnder interferometer
has the property that there exists only forward scattering,
\emph{i.e.} every incoming particle is transmitted in one of two
output arms. As a consequence the transmission sub-matrix is a
unitary matrix itself. It follows that
$\S_{41}\S^\dag_{13}+\S_{42}\S^\dag_{23}=0$. Here $1$ and $2$ are
contacts under ac-excitation and $3$ and $4$ are measurement
contacts. As a consequence the HBT-phase is given by $\chi=\pi$.
In particular there are no correlations if two in-phase voltages
are applied to two input contacts of a Mach-Zehder interferometer.

We now investigate the {\it statistical} properties of current
cross-correlations and HBT-phase in the chaotic quantum dots,
connected to four single channel leads. Chaotic scattering inside
the dot leads to substantial back-scattering and we expect
therefore nontrivial HBT-phase behavior. Scattering properties of
an open quantum dot are very sensitive to external conditions such
as shape (which can be changed by gate voltages), impurity
distribution or applied magnetic fields. This allows one to
explore the total ensemble of quantum dots of a proper symmetry
(presence or absence of time-reversal symmetry, TRS). Current
cross-correlations and HBT-phase are complicated functions of
transmission amplitudes and phases of the scattering matrix
elements, so the distributions can be obtained only by numerical
integration.

\begin{figure}
\includegraphics[height=6cm]{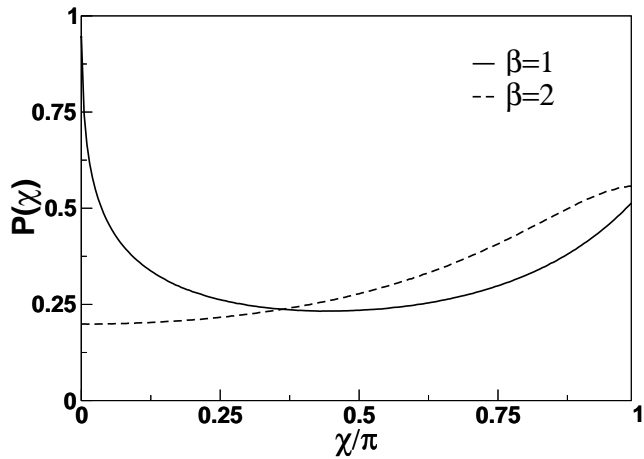}
\caption{Mesoscopic distribution $P_\b(\chi)$ of HBT-phase $\chi$
that maximizes the value of cross-correlations $S_{34}$ in quantum
dots. Angle $\chi$ is normalized by $\pi$ and
$P_\b(-\chi)=P_\b(\chi)$. The difference between the presence of
TRS (solid)
 and absence of TRS (dashed) is due to different distributions $P_\b({\cal T})$ in Eq. (\ref{eq:polar})
} \label{fig:phase}
\end{figure}

Using convenient
'polar decomposition' of the matrix $\cal S$:
\begin{eqnarray}\label{eq:polar}
{\cal S}=\left(%
\begin{array}{cc}
  u' & 0 \\
  0 & v \\
\end{array}%
\right)\left(%
\begin{array}{cc}
  \sqrt{1-{\cal T}}  & i\sqrt{{\cal T}} \\
  i\sqrt{{\cal T}} & \sqrt{1-{\cal T}}  \\
\end{array}%
\right)
\left(%
\begin{array}{cc}
  u & 0 \\
  0 & v' \\
\end{array}%
\right),
\end{eqnarray}
where $u,v,u',v'$ are unitary $2\times 2$ block matrices
($u'=u^{\rm T}, v'=v^{\rm T}$ in time-reversal symmetric case) and
${\cal T}=\mbox{diag}({T_1,T_2})$ with $T_i\in[0,1]$ distributed
according to the relevant symmetry class $\beta=1(2)$ with
(without of) the TRS,\cite{BeeReview} we express the noise
correlation $S$ of Eq. (\ref{eq:cross_corr_phase_shift}) in the
form,
\begin{eqnarray}\label{eq:HBTcorr}
S(\Delta\phi)=-\frac{e^2\o}{2\pi}{\cal P}|(v\sqrt{{\cal
T}}u\exp(i\sigma_z\frac{\Delta \phi}{2})u^\dagger\sqrt{{\cal T}}
v^\dagger)_{12}|^2.
\end{eqnarray}
Numerical integration over random matrix ensembles gives the
mesoscopic distribution $P_\b(\chi)$, $\b=1,2$ of the HBT-phase
Fig. \ref{fig:phase} presents the results for these distributions
$P_{\b}(\chi)$ in the range $0\leq\chi\leq \pi$, since they are
symmetric with respect to $\chi\to -\chi$. The distribution
$P_1(\chi)$ has peaks at $\chi=0,\pi$, and the distribution
$P_2(\chi)$ is monotonic for $0\leq\chi\leq \pi$. Below we present
a qualitative explanation of these differences.

The matrices $u,v$ are randomly distributed over the unitary
group, so the distributions $P_{\b}(\chi)$ differ only because of
different joint distributions of the transmission eigenvalues,
$P_\b(T_1,T_2)$. If one analyzes the distribution $P_\b(\chi)$,
two simple limits are readily calculated. First, take the limit
$T_1=T_2$ (unreachable for chaotic quantum dots, since the joint
distribution is $P_\b(T_1,T_2)\propto |T_1-T_2|^\beta(T_1
T_2)^{-1+\beta/2}$, but still instructive). It is easy to see that
 in this case $\chi=\pi$. One would expect that for sufficiently
close $T_1,T_2$ the phase $\chi$ does not differ much from
$\chi\approx \pi$ (for the Mach-Zender interferometer $T_1 =
T_2=1$ and the HBT-phase $\chi$ is locked at $\pi$). The relative
statistical weight of this limit in the mesoscopic distribution
$P_{\b}(\chi)$ might be characterized by the average of $\sqrt{T_1
T_2}/|T_1-T_2|$, this average is 3/4 for $\beta=1$ and 4/5 at
$\beta=2$. This could explain why $P_1(\pi)$ is slightly smaller
then $P_2(\pi)$.

Another example which provides some understanding is the case of
$T_1=0$ and arbitrary $T_2$. At $\Delta\phi=0$ the noise $S$
reaches its maximal value $|S^+|=e^2\o{\cal P}/\pi$, so that
$\chi=0$. The statistical weight of such cases could be
characterized by the average of $\sqrt{T_1/T_2}$ which is
divergent for $\b=1$ and finite for $\b=2$. Thus for $\b=1$ a peak
at $\chi=0$ could be expected.
The monotonic almost uniform $P_2(\chi)$ is hard to explain. We
expect that in the multi-channel limit $N\to\infty$, when the
transmission value distribution reaches its symmetry-insensitive
shape $P(T)\propto 1/\sqrt{T(1-T)}$, the differences between
$\b=1,2$ are washed out and the distributions $P_{1(2)}(\chi)$
become uniform. Possibly, the distribution $P_2(\chi)$ is much
more uniform then $P_1(\chi)$ in the four-mode quantum dot because
the $\cal S$ matrix is characterized by a much larger number of
independent variables (16 for $\b=2$ vs. 10 for $\b=1$). The
complicated behavior of the distribution of the HBT-phase
$P_\b(\chi)$ can be contrasted with that of conductance or
concurrence,\cite{bk} which depend only on transmission
eigenvalues of channels.

\begin{figure}
\includegraphics[height=6cm]{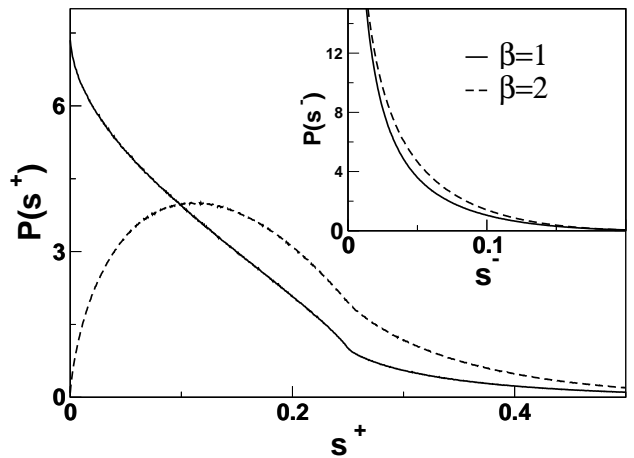}
\caption{Distribution functions of $S^+$ and $S^-$ (inset) for
quantum dots with (solid) and without (dashed) time-reversal
symmetry.
 Mesoscopic averages are similar in both cases, but the fluctuations are large.
 At small arguments $S^-$, the distributions on the inset are divergent as $P(S^-)\propto 1/\sqrt{S^-}$.}
\label{fig:concurrence}
\end{figure}

We next consider the mesoscopic distribution of the maximal
(minimal) values of the noise correlations $S$, which could be
reached in an experiment by tuning the phase shift $\Delta\phi$
between applied voltages. Using numerical integration, we find a
distribution of $S^+$ (main figure in Fig. \ref{fig:concurrence})
and $S^-$ (shown in the inset), normalized by a factor $e^2\o{\cal
P}/2\pi$. From the numerical integration we conclude that the
distribution of the minimal values $S^-$ diverges at small
arguments as $P(S^-)\propto (S^-)^{-1/2}$ for both $\b=1,2$. The
monotonic distributions of $S^-$ displayed in the inset quickly
decay, and the averaged values of $S_\b^-$ are
$S_1^-=0.022,S_2^-=0.027$. The distributions of $S^+$ are broad,
as expected from mesoscopic distributions in few-channel systems,
and the averaged values are also comparable,
$S_1^+=0.121,S_2^+=0.173$.

%-------------------------------------------------------------
\section{Conclusions}\label{sec:conclusions}
We constructed the scattering theory for electron-hole transport
in mesoscopic systems photon-excited at contacts. In the limit of
a weak ac-excitation we investigate the correlation between
electrons and holes. In different geometries signatures of such
correlations appear in a different way. In a two terminal
mesoscopic conductor subjected to a weak ac-voltage at one of the
contacts, the electron-hole correlation effect in the shot noise
co-exists with electron-electron correlations. In a three terminal
geometry we investigate the correlations between currents at
different terminals and find that they are pure electron-hole
correlations.

In contrast to dc-biased systems, in the ac-biased systems
investigated here we can vary not only the magnitude of the
applied voltages but also the phases of the applied voltages. This
provides additional controllable parameters. Shot noise
measurements in a four terminal mesoscopic conductor provide
information about two particle exchange interference in the
sample. The shot noise depends on the relative phase between
applied voltages, because the particles from different sources are
indistinguishable.

We illustrate our theory on the example of a four probe chaotic
dot coupled to four single channel leads. At two leads the
conductor is subject to ac-voltages. When the phase shift
$\Delta\phi$ of the applied voltages coincides with the HBT-phase
$\chi$ of the sample, the correlations reach their maximal value
$S^+$. One might expect the phase to be uniformly distributed.
However the quantum dot coupled to single channel leads exhibits a
strongly non-uniform mesoscopic distributions $P(\chi)$ of the
HBT-phase and of $P(S^\pm)$ of the extremal correlations values
$S^\pm$. Recent advances in high-frequency measurement
 techniques\cite{glattli,schoel,kozh} will make it possible to
measure these distributions. The close link between the
two-particle HBT-effect \cite{sam2} and quasi-particle
entanglement \cite{bee1,sam2} and recent proposals for dynamic
generation of quasi-particle entanglement \cite{sam3,bee2,llb}
make such experiments highly desirable.

\section*{ACKNOWLEGEMENTS}
 This work was supported
by the Swiss National Science Foundation and by the  M. Curie RTN
on "Fundamentals of Nanoelectronics".

\appendix
\section{Quantum state and Exchange Interference}\label{sec:exchange}

In this appendix we construct the quantum state which leads to the
Hanbury Brown Twiss exchange interference effect discussed in Sec.
\ref{sec:phase}. We consider the conductor shown in Fig.
\ref{fig3} with four one channel leads.  In particular we want to
show that only electron hole pairs which are created in either
lead $1$ or $2$ and are subsequently split into lead $3$ or $4$
generate the HBT-exchange interference term.

Each state incident from a reservoir subject to an oscillating
voltage can gain or loose modulation energy quanta. The many
particle state incident from the $\a$-th reservoir,
$\prod_{\epsilon<0}{a}'^\dag_\a(\epsilon)|\rangle$, can be
transformed using Eq. (\ref{eq:operator_connection}) into the
state $|in\rangle$ incident on the mesoscopic conductor.
$|\rangle$ is the true vacuum. Taking only the first sidebands
into account, we find
\begin{equation}\label{eq:res_state}
|in\rangle=\prod_{E<0}\left({a}_\a^\dag(\epsilon)+V_\a^{+}{a}_\a^\dag(\epsilon^+)+V_\a^{-}{a}_\a^\dag(\epsilon^-)
\right)|\rangle.
\end{equation}
Here $V^{\pm}_\a=eV_\a e^{\pm i\phi_\a}/2\h\o$ is the probability
amplitude for the creation of an electron-hole pair in the $\a$-th
lead and $\epsilon^\pm=\epsilon\pm\h\o$. Using the commutation
rules of free fermion operators we find the state incident on the
four-terminal mesoscopic conductor in the presence of oscillating
potentials at contacts $1$ and $2$:
\begin{eqnarray}\label{eq:in_state}
|in\rangle=|0\rangle + \int_{0}^{\h\o} d\epsilon {a}^\dag(\epsilon)\V
{a}(\epsilon^-)|0\rangle\,.
\end{eqnarray}
Here the vacuum state
$|0\rangle=\prod_{(\epsilon<0,\a=\overline{1,4})}{a}^\dag_\a(\epsilon)|\rangle$
is the filled Fermi sea at equilibrium in all four leads. The
second term on the r.h.s. of Eq. (\ref{eq:in_state}) represents
the superposition of the electron-hole pairs coming from leads $1$
and $2$. The matrix $\V$ is diagonal $\V={\mbox{diag}}
(eV_1e^{i\Delta\phi}/2\h\o,eV_2/2\h\o,0,0)$. Using the relation
between incoming and outgoing states $a_\a=\sum_\b\S^\dag_{\a\b}
b_\b$ we find for the outgoing state:
\begin{eqnarray}\label{eq:out_state}
|out\rangle=|\bar 0\rangle+\int^{\h\o}_0 d\epsilon
{b}^\dag(\epsilon)\S\V\S^\dag {b}(\epsilon^-)|\bar0\rangle\,.
\end{eqnarray}
Here the new vacuum $|\bar
0\rangle=\prod_{(\epsilon<0,\a=\overline{1,4})}\det(\S){b}^\dag_\a(\epsilon)|\rangle$
describes the equilibrium state in the basis of out-going states.

The many-particle state (\ref{eq:out_state}) contains information
about the final states in all the leads. We are specifically
interested in the correlation of currents at leads $3$ and $4$ for
a conductor subject to oscillating potentials at contacts $1$ and
$2$. Leads $3$ and $4$ are grounded and at zero temperature there
are no particles injected into the mesoscopic conductor through
these leads. Thus the cross-correlations $S_{34}$ are determined
only by that portion of the state (\ref{eq:out_state}) which
describes out-going particles in lead $3$ and $4$,
\begin{eqnarray}
\label{eq:out34}
|out \rangle_{34}&=&\frac{eV}{2\h\o}\int^{\h\o}_0
d\epsilon
\left[...\right]|\bar{0}\rangle, \\
\left[...\right]&=&\left(e^{i\Delta\phi}\S_{41}\S^\dag_{13}+\S_{42}\S^\dag_{23}\right)
b_4^\dag(\epsilon)b_3(\epsilon^-)\nonumber \\ &+&
\left(e^{i\Delta\phi}\S_{31}\S^\dag_{14}+\S_{32}\S^\dag_{24}\right)
b_3^\dag(\epsilon)b_4(\epsilon^-).
\end{eqnarray}

An excitation with energy above zero corresponds to an electron
$b_\a^\dag(\epsilon)=e_\a^\dag(\epsilon)$ and an excitation with
energy below zero to a hole
$b_\a(\epsilon^-)=h_\a^\dag(\epsilon^-)$. Thus the first term in
Eq. (\ref{eq:out34}) represents a superposition of amplitudes for
an electron-hole pair created in contact $1$ or contact $2$ with
the electron leaving through  $4$ and the hole through contact
$3$. It is an orbitally entangled electron-hole pair state and the
index of the source contact $1$ and $2$ is a pseudo-spin index.
Similarly the second term represents orbitally entangled
electron-hole pairs with the electron leaving through contact $3$
and the electron through contact $4$. In fact, from a formal point
of view the state is identical to the one created by two
oscillating potentials acting in spatially separated interior
regions of a conductor investigated in Ref. [\onlinecite{sam3}].
One might think that the oscillating potentials in contacts $1$
and $2$ serve to mark particles created in these contacts, since
they carry the phase factor $\exp(-i \phi_1 )$ and $\exp(-i \phi_2
)$ and thus to make them distinguishable. However, also for the
applied oscillating voltages it is only the phase difference
$\Delta \phi = \phi_1 - \phi_2$ that counts. As a consequence, the
phase of the oscillating voltages only modulates the
HBT-interference but does not destroy it.

We now show how the different terms in the quantum state Eq.
(\ref{eq:out34}) contribute to the HBT-current-correlation between
leads $3$ and $4$. This correlation can now be derived by using
either the full state (\ref{eq:out_state}) or the state
(\ref{eq:out34}) which contains the out-going particles in leads
$3$ and $4$ only. The results of these two calculations are of
course identical. Using Eq. (\ref{shot_noise_step3}) we find:
\begin{eqnarray} \label{eq:snapp1} S_{34}&=&2\pi\h\int\int d\e
d\e' \langle out|\Delta I_3(\e)\Delta
I_4(\e')|out\rangle\nonumber\,.\end{eqnarray}
Since leads $3$ and $4$ are grounded and there are no particles
injected into the
mesoscopic conductor through these leads,
the current operators
in $3$ and $4$ can be written in terms of operators of the outgoing
states only:
$$I_\a(\e)=-\frac{e}{2\pi\h}{b}_\a^\dag(\e){b}_\a(\e).$$
For weak ac-potentials, the fact that average currents are equal
to zero to linear order in the potentials, permits us to write:

\begin{widetext}
\begin{eqnarray} \label{eq:snapp2} S_{34}=\frac{e^2}{h}\int d\e
d\e'\int_{0}^{\h\o} d\epsilon d\epsilon'
\sum_{\substack{\a\b\g\d}}\left[\S \V^\dag \S^\dag
\right]_{\a\b}\left[\S  \V \S^\dag \right]_{\g\d} \langle\bar 0|{
b}^\dag_\a(\epsilon^-){ b}_\b(\epsilon){ b}^\dag_3(\e){ b}_3(\e){
b}^\dag_4(\e'){ b}_4(\e'){ b}^\dag_\g(\epsilon'){
b}_\d(\epsilon'^-)|\bar 0\rangle . \end{eqnarray}
\end{widetext}
Since $\h\o>\epsilon>0$, the operators
${b}^\dag_\a(\epsilon-\h\o)$ and ${b}_\b(\epsilon)$ acting to the
right on the vacuum give zero. Applying Wick's theorem to the
quantum mechanical average in Eq. (\ref{eq:snapp2}) and performing
the integration over the energies we find:

\begin{eqnarray} \label{eq:snapp3}
S_{34}&=&-\frac{e^2\o}{2\pi} \left|\left(\S \V
\S^\dag\right)_{34}\right|^2 \,.
\end{eqnarray}

This expression coincides with Eq.
(\ref{eq:cross_corr_phase_shift}). The diagonal matrix $\V$
provides additional selection rules for the processes which
contribute to the shot noise. Denoting the part of $\V$ in lead
$\a$ by $\V_\a$ we find from Eq. (\ref{eq:snapp3})
\begin{eqnarray}
\left|\left(\S \V \S^\dag\right)_{34}\right|^2&=& \left(\S_{31} \V^\dag_{1} \S^\dag_{14}+\S_{32} \V^\dag_{2} \S^\dag_{24}\right)\nonumber \\
&&\times\left(\S_{41} \V_{1} \S^\dag_{13}+\S_{42} \V_{2}
\S^\dag_{23}\right).
\end{eqnarray}

Thus we see that of the four terms in the current correlation
function, two correspond to electron-hole emission out of the same
contact, and two contributions, sensitive to the phase $\Delta
\phi$, correspond to electron-hole emission from both contacts $1$
and $2$. All contributions arise due to electron-hole pairs which
are split into contact $3$ and $4$.
\section{ Current correlations for Poissonian source of $e-h$ pairs}\label{sec:Poisson}

Here we present a model of a probability game to illustrate
correlations between photon-generated $e-h$ pairs similarly to the
probabilistic electron model of Ref. [\onlinecite{binomial}]. We
calculate correlations of electrons and holes for a two-terminal
geometry, and show that the total noise can not be interpreted in
terms of {\it independent} electron and hole noises.

Consider a two-lead geometry with one of the leads exposed to a
monochromatic photon beam of frequency $\omega$. The photons
generate $e-h$ pairs, and the statistics of these pairs is assumed
to be Poissonian. (The details of the photon statistics are quite
irrelevant and we take this example as an illustration.) Let $\l$
be the average number of pairs generated in a period $2\pi/\o$.
The number $N$ of pairs generated during one period fluctuates
according to the Poissonian distribution $
\rho_\l(N)=\l^N\exp(-\l)/N!$. A barrier of transparency ${\cal T}$
separates the left and right contacts of the conductor. The
probability that $m$ electrons and $n$ holes out of $N$
electron-hole pairs are transmitted from the left to the right
contact is
\begin{eqnarray}
P(m,n|N)=\binom{m}N\binom{n}N {\cal T}^{m+n}(1-{\cal
T})^{2N-(m+n)}.
\end{eqnarray}
The measured quantity is the {\it charge} current $j=m-n$ (we omit
electron charge $e$ for simplicity) and its distribution is
symmetric with respect to $j=0$, so to be definite we consider
$j>0$. The distribution $P(j)$ is then given by
\begin{eqnarray}\label{eq:P}
P(j)= \sum_{N=1}^\infty \sum_{m=j}^N P(m,m-j|N)\rho_\l(N).
\end{eqnarray}
The limits of summations result from the fact that to create the
current $j$ one needs at least $j$ electrons out of $N\geq m$
electron-hole pairs. For $\l\ll 1$ which we assume from now on, the
leading contribution to Eq. (\ref{eq:P}) is
\begin{eqnarray}\label{eq:P_eh_Poisson}
P(j)&\approx& \frac{(\l {\cal T}(1-{\cal T}))^j}{j!},
\end{eqnarray}
and thus $\langle j^{2k+1}\rangle=0$, $\langle
j^{2k}\rangle\approx 2\l {\cal T}(1-{\cal T})$. We conclude that
the noise equals $2\l {\cal T}(1-{\cal T})$. From comparison with
the scattering theory, summing up the Eqs. (\ref{eq:2term_corr}),
we find $\langle j^2\rangle=(eV/2\hbar\o)^2{\cal T}(1-{\cal T})$,
which allows us to identify $\lambda=(eV)^2/8(\h\o)^2\ll 1$.

 If we now consider the distribution of the
electron current $j = j_e$ and hole current $j = j_h$ separately,
we find
\begin{eqnarray}\label{eq:P_e_Poisson}
P(j)&=&\sum_{N=j}^\infty \frac{N!\rho_\l(N)}{j!(N-j)!} {\cal
T}^{j}(1-{\cal T})^{N-j} =\rho_{\l{\cal T}}(j). \nonumber
\end{eqnarray}
The distribution of the current $j$ in the r.h.s. is a Poissonian
distribution characterized by the parameter $\l{\cal T}$. However,
since now the total measured current is the difference between two
{\it independent} Poissonian processes, it is not a Poissonian
process itself (only for two independent Gaussian processes is
their difference also a Gaussian process). For $j=j_e-j_h$ we find
\begin{eqnarray}
P(j)= I_{|j|}(2\l {\cal T})\exp(-2\l {\cal T}),
\end{eqnarray}
with $I_{|j|}(x)$ the modified Bessel function of the $|j|$th
(integer) order. Thus we find $\langle j^2\rangle= 2\l {\cal
T}+{\cal O}((\l{\cal T})^2)$, unlike the correct result $2\l {\cal
T}(1-{\cal T})$ found from the distribution Eq.
(\ref{eq:P_eh_Poisson}).

\end{document}